\documentclass[aps,reprint,floatfix,superscriptaddress,amsmath,amssymb,pra]{revtex4-1}

\usepackage{graphicx}

\newcommand{\be}{\begin{equation}}
\newcommand{\ee}{\end{equation}}

\newcommand{\ket} [1] {| #1 \rangle}

\newcommand{\ttr}{\mathrm{tTr}}
\newcommand{\tbf}[1]{\textbf{#1}}

\begin{document}
\title{Tensor network decompositions in the presence of a global symmetry}
\author{Sukhwinder Singh}
\author{Robert N. C. Pfeifer}
\author{Guifr\'e Vidal}
\affiliation{The University of Queensland, Department of Physics, Brisbane, QLD 4072, Australia}

\begin{abstract}
Tensor network decompositions offer an efficient description of certain many-body states of a lattice system and are the basis of a wealth of numerical simulation algorithms. We discuss how to incorporate a global symmetry, given by a compact, completely reducible group $\mathcal{G}$, in tensor network decompositions and algorithms. This is achieved by considering tensors that are invariant under the action of the group $\mathcal{G}$. Each symmetric tensor decomposes into two types of tensors: \emph{degeneracy tensors}, containing all the degrees of freedom, and \emph{structural tensors}, which only depend on the symmetry group. In numerical calculations, the use of symmetric tensors ensures the preservation of the symmetry, allows selection of a specific symmetry sector, and significantly reduces computational costs. On the other hand, the resulting tensor network can be interpreted as a superposition of exponentially many \emph{spin networks}. Spin networks are used extensively in loop quantum gravity, where they represent states of quantum geometry. Our work highlights their importance also in the context of tensor network algorithms, thus setting the stage for cross-fertilization between these two areas of research.
\end{abstract}


\maketitle

\emph{Locality} and \emph{symmetry} are pivotal concepts in the formulation of physical theories. In a quantum many-body system, locality implies that the dynamics are governed by a Hamiltonian $H$ that decomposes as the sum of terms involving only a small number of particles, and whose strength decays with the distance between the particles. In turn, a symmetry of the Hamiltonian $H$ allows us to organize the kinematic space of the theory according to the irreducible representations of the symmetry group.

Both symmetry and locality can be exploited to obtain a more compact description of many-body states and to reduce computational costs in numerical simulations. In the case of symmetries, this has long been understood. Space symmetries, such as invariance under translations or rotations, as well as internal symmetries, such as particle number conservation or spin isotropy, divide the Hilbert space of the theory into sectors labeled by quantum numbers or charges. The Hamiltonian $H$ is by definition block-diagonal in these sectors. If, for instance, the ground state is known to have zero momentum, it can be obtained by just diagonalizing the (comparatively small) zero momentum block of $H$. 

In recent times, also the far-reaching implications of locality for our ability to describe many-body systems have started to unfold.
The local character of Hamiltonian $H$ limits the amount of entanglement that low energy states may have, and in a lattice system, restrictions on entanglement can be exploited to succinctly describe these states with a tensor network (TN) decomposition. 
Examples of TN decompositions include \emph{matrix product states} (MPS) \cite{MPS}, \emph{projected entangled-pair states} \cite{PEPS}, and the \emph{multi-scale entanglement renormalization ansatz} (MERA) \cite{MERA}. Importantly, in a lattice made of $N$ sites, where the Hilbert space dimension grows exponentially with $N$, TN decompositions often offer an efficient description (with costs that scale roughly as $N$). This allows for scalable simulations of quantum lattice systems, even in cases that are beyond the reach of standard Monte Carlo sampling techniques. As an example, the MERA has been recently used to investigate ground states of frustrated antiferromagnets \cite{MERAanti}.

In this paper we investigate how to incorporate a global symmetry into a TN, so as to be able to simultaneously exploit both the locality and the symmetries of physical Hamiltonians to describe many-body states. Specifically, in order to represent a symmetric state that has a limited amount of entanglement, we use a TN made of symmetric tensors. This leads to an \emph{approximate}, efficient decomposition that preserves the symmetry \emph{exactly}. Moreover, a more compressed description is obtained by breaking each symmetric tensor into several degeneracy tensors (containing all the degrees of freedom of the original tensor) and structural tensors (completely fixed by the symmetry). This decomposition leads to a substantial reduction in computational costs and reveals a connection between TN algorithms and the formalism of spin networks \cite{SpinNetwork} used in loop quantum gravity \cite{LQG}.

In the case of a MPS, global symmetries have already been studied by many authors (see e.g. \cite{MPS,symmMPS}) both in the context of 1D quantum systems and 2D classical systems. A MPS is a trivalent TN (i.e. each tensor has at most 3 indices) and symmetries are comparatively easy to characterize. The present analysis applies to the more challenging case of a generic TN decomposition (where tensors typically have more than three indices).

\begin{figure}[t]
  \includegraphics[width=8cm]{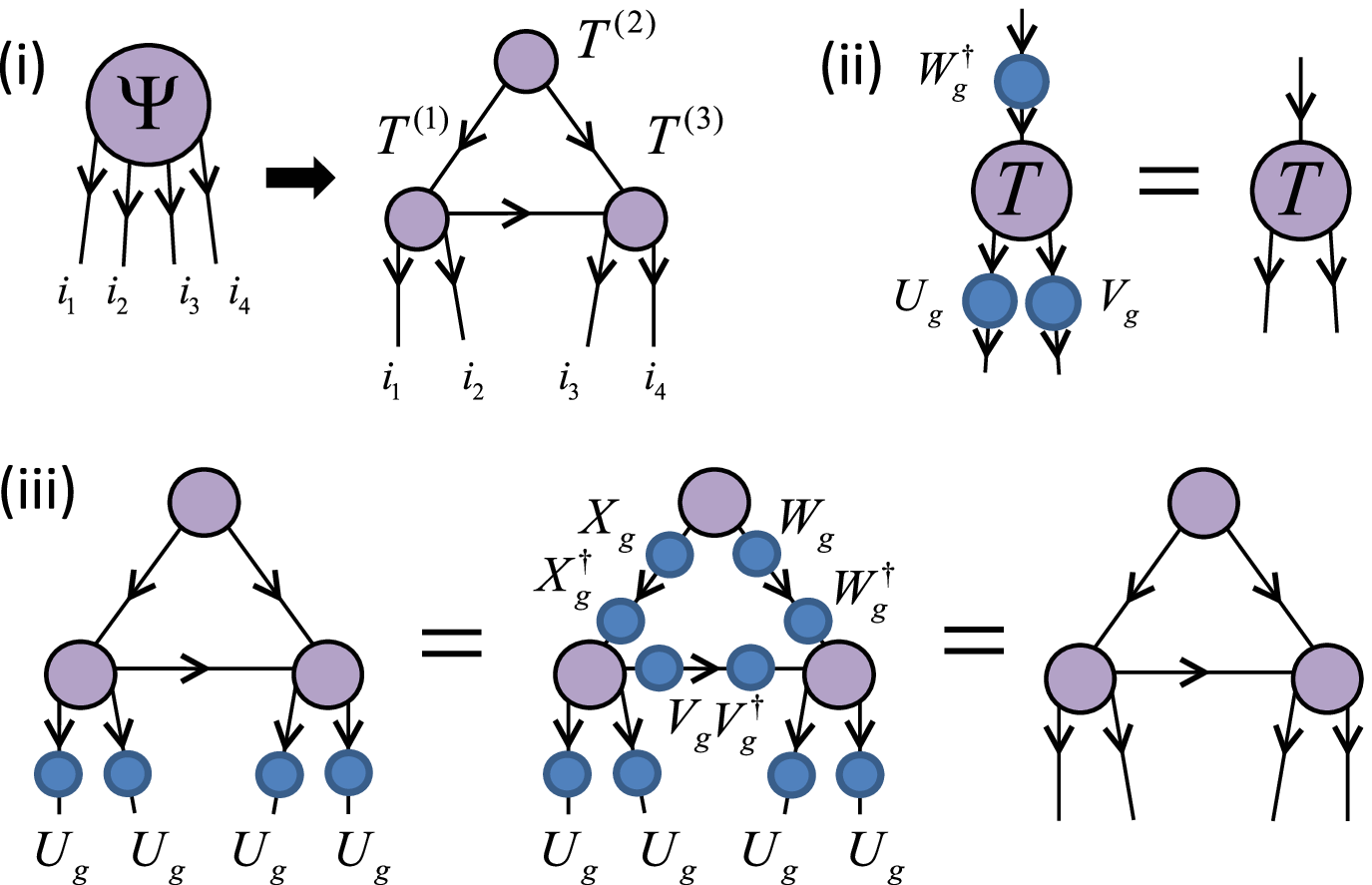}
\caption{(Color online) (i) Four-site state $\Psi$ expressed in terms of a tensor network made of three tensors connected according to a directed graph. (ii) Invariance of tensor $T$ in Eq. \ref{eq:symmT}. (iii) Invariance of a tensor network of symmetric tensors, Eq. \ref{eq:symm}.}
\label{fig:symmTN}
\end{figure}

We consider a lattice $\mathcal{L}$ made of $N$ sites, where each site is described by a complex vector space $\mathbb{V}$ of finite dimension $d$. A pure state $\ket{\Psi}\in\mathbb{V}^{\otimes N}$ of the lattice can be expanded as
\begin{equation}
	\ket{\Psi} = \sum_{i_1,i_2, \ldots, i_N=1}^d (\Psi)_{i_1i_2\ldots i_N} \ket{i_1 i_2 \ldots i_N}, 
\end{equation}
where $\ket{i_s}$ denotes a basis of $\mathbb{V}$ for site $s \in \mathcal{L}$.
For our purposes, a TN decomposition for $\ket{\Psi}$ consists of a set of tensors $T^{(v)}$ and a network pattern or graph characterized by a set of vertices and a set of directed edges. Each tensor $T^{(v)}$ sits at a vertex $v$ of the graph, and is connected with neighboring tensors by \emph{bond} indices according to the edges of the graph. The graph also contains $N$ open edges, corresponding to the $N$ \emph{physical} indices $i_1,i_2, \ldots, i_N$. The $d^N$ coefficients $(\Psi)_{i_1i_2\ldots i_N}$ are expressed as (Fig. \ref{fig:symmTN}.i)
\begin{equation}
	(\Psi)_{i_1 i_2 \ldots i_N} = \ttr \left(\bigotimes_{v} T^{(v)}\right),
	\label{eq:TN}
\end{equation}
namely as the tensor product of the tensors $T^{(v)}$ on all the vertices $v$, where the \emph{tensor trace} $\ttr$ contracts all bond indices, so that only the physical indices $i_1, i_2, \ldots i_N$ remain on the r.h.s. of Eq. \ref{eq:TN}.

We also introduce a compact, completely reducible group $\mathcal{G}$.  This includes finite groups as well as Lie groups such as O(n), SO(n), U(n), and SU(n). Let $U:\mathcal{G}\rightarrow L(\mathbb{V})$ be a unitary matrix representation of $\mathcal{G}$ on the space $\mathbb{V}$ of one site, so that for each $g\in \mathcal{G}$, $U_g:\mathbb{V} \rightarrow \mathbb{V}$ denotes a unitary matrix and $U_{g_1g_2} = U_{g_1}U_{g_2}$. Here we are interested in states $\ket{\Psi}$ that are invariant under transformations of the form $U_g^{\otimes N}$ \cite{covariant},
\begin{equation}
	(U_g)^{\otimes N} \ket{\Psi} = \ket{\Psi},~~~~~\forall~g\in \mathcal{G}.
	\label{eq:symm}
\end{equation}
The space $\mathbb{V}$ of one site decomposes as the direct sum of irreducible representations (irreps) of $\mathcal{G}$,
\begin{equation}
	 \mathbb{V} \cong \bigoplus_a d_a \mathbb{V}^{a} \cong  \bigoplus_a \left( \mathbb{D}^{a} \otimes \mathbb{V}^{a}\right), 
	 \label{eq:irreps}
\end{equation}
where $\mathbb{V}^{a}$ denotes the irrep labeled with charge $a$ and $d_a$ is the number of times $\mathbb{V}^{a}$ appears in $\mathbb{V}$. We denote by $a=0$ the charge corresponding to the trivial irrep, so that $\mathbb{V}^{0} \cong \mathbb{C}$ and $U^{a}_g=1$. In Eq. \ref{eq:irreps} we have also rewritten the same decomposition in terms of a $d_a$-dimensional degeneracy space $\mathbb{D}^{a}$. We choose a local basis $\ket{i} = \ket{a,\alpha_a,m_a}$ in $\mathbb{V}$, where $\alpha_a$ labels states within the degeneracy space $\mathbb{D}^{a}$ (i.e. $\alpha_a=1,\ldots,d_a$) and $m_{a}$ labels states within irrep $\mathbb{V}^{a}$. In this basis, $U_g$ reads
\begin{equation}
	U_g = \bigoplus_a \left( \mathbb{I}^{a} \otimes U_g^{a}\right).
\end{equation}
Recall that an operator $M:\mathbb{V}\rightarrow \mathbb{V}$ that commutes with the group, $[M,U_g]=0$ for all $g\in\mathcal{G}$, decomposes as \cite{group}
\begin{equation}
	M = \bigoplus_a \left( M^{a} \otimes \tilde{\mathbb{I}}^{a}\right)~~~~~~~~ \mbox{Schur's lemma.}
	\label{eq:Schur}
\end{equation}

\begin{figure}[t]
  \includegraphics[width=8cm]{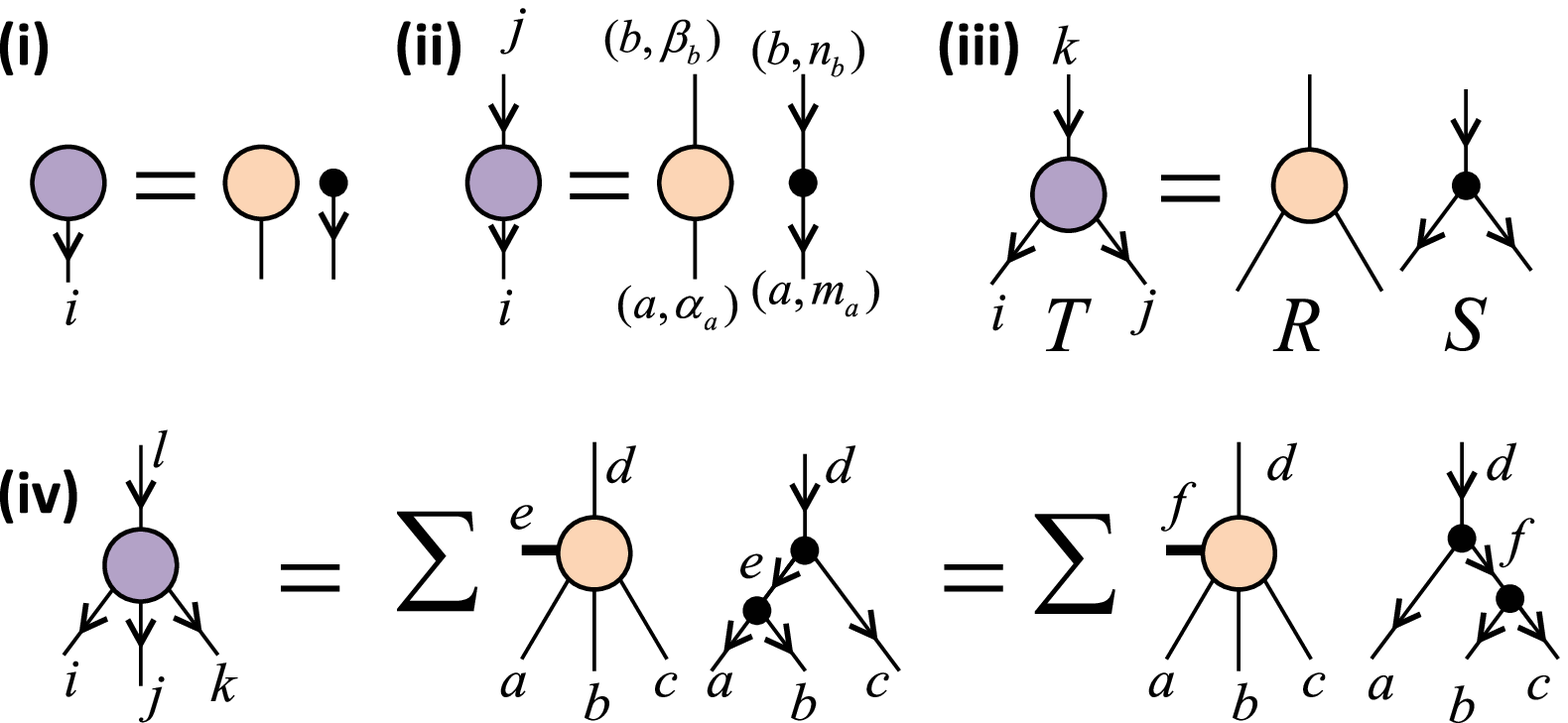}
\caption{(Color online) Decomposition of tensors with one to four indices. The sums in (iv) run over the intermediate indices $(e, \epsilon_e, q_e)$ and $(f,\zeta_f,r_f)$ in Eqs.~\ref{eq:T4}, \ref{eq:T4b}.}
\label{fig:T1234}
\end{figure}

Our goal is to characterize a TN made of symmetric tensors, namely tensors that are invariant under the simultaneous action of $\mathcal{G}$ on all their indices. A symmetric tensor $T$ with e.g. two outgoing indices $i$ and $j$ and one incoming index $k$ fulfills (Fig. \ref{fig:symmTN}.ii)
\begin{equation}
	\sum_{ijk} (U_g)_{i'i} (V_g)_{j'j}  (T)_{ijk} (W^{\dagger}_g)_{kk'}= (T)_{i'j'k'},  ~~ \forall\,g\in \mathcal{G},
	\label{eq:symmT}
\end{equation}
where $U$, $V$, and $W$ denote unitary matrix representations of $\mathcal{G}$. Clearly, this choice guarantees that Eq. \ref{eq:symm} is satisfied (Fig. \ref{fig:symmTN}.iii). Standard group representation theory results \cite{group} imply that each symmetric tensor can be further decomposed in such a way that the degrees of freedom which are not fixed by the symmetry can be isolated (Fig. \ref{fig:T1234}). Next we discuss the cases of 
tensors with a 
small number of indices. Recall that an index $i$ of a tensor is associated with a vector space that decomposes as in Eq. \ref{eq:irreps}, and therefore we can write $i = (a,\alpha_a,m_a)$, $j = (b, \beta_b, n_b)$, $k=(c, \gamma_c, o_c)$ and so on.

%%%%%%%%%%%%%%%%%%%%%%%%%%%%%%%%%%%%%
%%%%%%%%%%%%%%%%%%%%%%%%%%%%%%%%%%%%%
%%%%%%%%%%%%%%%%%%%%%%%%%%%%%%%%%%%%%
%%%%%%%%%%%%%%%%%%%%%%%%%%%%%%%%%%%%%

\emph{One leg.---} A tensor $T$ with only one index $i$ is invariant only if $\mathcal{G}$ acts on it trivially, so %that 
the only relevant irrep is $a=0$, and index $i=\alpha_{0}$ labels states within the degeneracy space $\mathbb{V}^{0}$. 

\emph{Two legs.---} Schur's lemma \cite{group} establishes that a symmetric tensor $T$ with one outgoing index $i$ and one incoming index $j$ decomposes as (cf. Eq. \ref{eq:Schur})
\begin{equation}
	(T)_{ij} = (P^{ab})_{\alpha_a \beta_b} (Q^{ab})_{m_an_b}, ~~~~Q^{ab}= \delta_{ab} \delta_{m_a n_b}.
\label{eq:T2}
\end{equation}
Thus, for fixed values of the charges $a$ and $b$, $(T)_{ij}$ breaks into a \emph{degeneracy tensor} $P^{ab}$ (where only $a=b$ is relevant) and another tensor $Q^{ab}$. $P^{ab}$ contains all the degrees of freedom of $T$ that are not fixed by the symmetry, whereas $Q^{ab}$ is completely determined by $\mathcal{G}$. 
Another combination of outgoing and incoming indices, say two incoming indices, leads to a different form for tensor $Q^{ab}$.

\emph{Three legs.---} The tensor product of two irreps with charges $a$ and $b$ can be decomposed as the direct sum of irreps
\begin{equation}
	\mathbb{V}^{a}\otimes \mathbb{V}^{b} \cong \bigoplus_{c} N_{ab}^c \mathbb{V}^{c} ,
\end{equation}
where $N_{ab}^c$ denotes the number of copies of $\mathbb{V}^{c}$ that appear in the tensor product. For notational simplicity, from now on we assume that $\mathcal{G}$ is multiplicity free \cite{degeneracy}, i.e. $N_{ab}^c\leq 1$, and denote by $(Q^{abc})_{m_a n_b o_c}$ the change of basis between the product basis $\ket{a,m_{a}}\otimes\ket{b,n_b}$ and the coupled basis $\ket{c,o_c}$. Then Wigner-Eckart theorem states that a symmetric tensor $T$ with e.g. two outgoing indices $i,j$ and one incoming index $k$ decomposes as 
\begin{equation}
	(T)_{ijk} = (P^{abc})_{\alpha_a \beta_b \gamma_c} (Q^{abc})_{m_an_bo_c}.
	\label{eq:T3}
\end{equation}
As before, for fixed values of the charges $a,b,c$, $(T)_{ijk}$ factorizes into degeneracy tensors $P^{abc}$ with all the degrees of freedom and structural tensors $Q^{abc}$ (the Clebsch-Gordan coefficients) completely determined by the group $\mathcal{G}$. An analogous decomposition with different $Q^{abc}$ holds for other combinations of incoming and outgoing indices. 

%%%%%%%%%%%%%%%%%%%%%%%%%%%%%%%%%%%%%
%%%%%%%%%%%%%%%%%%%%%%%%%%%%%%%%%%%%%
%%%%%%%%%%%%%%%%%%%%%%%%%%%%%%%%%%%%%
%%%%%%%%%%%%%%%%%%%%%%%%%%%%%%%%%%%%%

\emph{Four legs.---}The tensor product of three irreps $\mathbb{V}^{a}\otimes \mathbb{V}^b \otimes \mathbb{V}^c $ may contain several copies of an irrep $\mathbb{V}^d$. Let $e$ be the charge that results from fusing $a$ and $b$, $\mathbb{V}^{a}\otimes \mathbb{V}^b = \bigoplus_e N_{ab}^{e} \mathbb{V}^e$. We can use the values of $e$ for which $N_{ab}^{e}N_{ec}^d \neq 0$ (i.e., such that $a$ and $b$ fuse to $e$; and $e$ and $c$ fuse to $d$) to label the different copies of $\mathbb{V}^d$ that appear in $\mathbb{V}^{a}\otimes \mathbb{V}^b \otimes \mathbb{V}^c$. Let $(Q^{abcd}_e)_{m_an_bo_cp_dq_e}$ denote the change of basis between the product basis $\ket{a m_a} \otimes \ket{b n_b} \otimes \ket{c o_c}$ and the coupled basis $\ket{d p_d; e}$ obtained by fusing to the intermediate basis $\ket{eq_e} \in \mathbb{V}^e$. Then a symmetric tensor $T$ with three outgoing indices $i$, $j$ and $k$ and one incoming index $l = (d,\delta_d,p_d)$ decomposes as
\begin{equation}
	(T)_{ijkl} = \sum_{e,\epsilon_e,q_e}(P^{abcd}_e)_{\alpha_a \beta_b \gamma_c \delta_d \epsilon_e} (Q^{abcd}_e)_{m_an_bo_cp_dq_e},
\label{eq:T4}
\end{equation}
where the sum is over all relevant values of the intermediate indices $(e, \epsilon_e, q_e)$. Alternatively, $T$ can be decomposed as
\begin{equation}
	(T)_{ijkl} = \sum_{f,\zeta_f,r_f}(\tilde{P}^{abcd}_f)_{\alpha_a \beta_b \gamma_c \delta_d\zeta_f} (\tilde{Q}^{abcd}_f)_{m_an_bo_cp_dr_f},
\label{eq:T4b}
\end{equation}
where $(\tilde{Q}^{abcd}_f)_{m_an_bo_cp_d}$ denotes the change of basis to another coupled basis $\ket{d p_d; f}$ of $\mathbb{V}^d$ obtained by fusing first $b$ and $c$ into $f$, and then $a$ and $f$ into $d$, involving a different set of intermediate indices $(f,\zeta_f,r_f)$. The two coupled bases are related by a unitary transformation given by the 6-index tensor $F$ [e.g. the 6-$j$ symbols for $\mathcal{G} =\mathrm{SU(2)}$] such that
\begin{equation}
	\tilde{Q}^{abcd}_f = \sum_{e} (F^{abc}_d)^{e}_{f} Q^{abcd}_{e}.
	\label{eq:Fmove}
\end{equation}
Since Eqs. \ref{eq:T4} and \ref{eq:T4b} represent the same tensor $T$, the degeneracy tensors $P$ and $\tilde{P}$ are related by
\begin{equation}
	\tilde{P}^{abcd}_f = \sum_{e} ({F^{abc}_d}^{*})^{e}_{f} P^{abcd}_{e}.
	\label{eq:Fmove2}
\end{equation}

More generally, a symmetric tensor $T$ with $t$ indices $i_{s}=(a_s,\alpha_{a_s},m_{a_s})$, where $s=1,\ldots, t$, decomposes as
\begin{eqnarray}
	(T)_{i_1 i_2 \ldots i_t} = \sum (P^{a_1 \ldots a_t}_{e_1 \ldots e_{t'}})_{\alpha_{a_1} \ldots \alpha_{a_t}, \alpha_{e_1}  \ldots \alpha_{e_{t'}}} \nonumber\\
	\times (Q^{a_1\ldots a_t}_{e_1 \ldots e_{t'}})_{m_{a_1} \ldots m_{a_t},m_{e_1} \ldots m_{e_{t'}}},
	\label{eq:decoT}
\end{eqnarray}
where the sum is over the intermediate indices $(e_k,\alpha_{e_k},m_{e_k}),~~k=1,\ldots, t'$. The degeneracy tensors $P^{a_1 \ldots a_t}_{e_1 \ldots e_{t'}}$ contain all the degrees of freedom of $T$, whereas the structural tensors $Q^{a_1 \ldots a_t}_{e_1 \ldots e_{t'}}$ are completely determined by the symmetry. Here $e_{1}, e_{2},\ldots, e_{t'}$ are intermediate charges that decorate the inner branches of a trivalent tree used to label a basis in the space of intertwining operators between the tensor products of incoming and outgoing irreps. A different choice of tree will produce different sets of tensors $\tilde{P}$ and $\tilde{Q}$, related to $P$ and $Q$ by F-moves \cite{abelian}.

%%%%%%%%%%%%%%%%%%%%%%%%%%%%%%%%%%%%%
%%%%%%%%%%%%%%%%%%%%%%%%%%%%%%%%%%%%%
\begin{figure}[t]
\includegraphics[width=8cm]{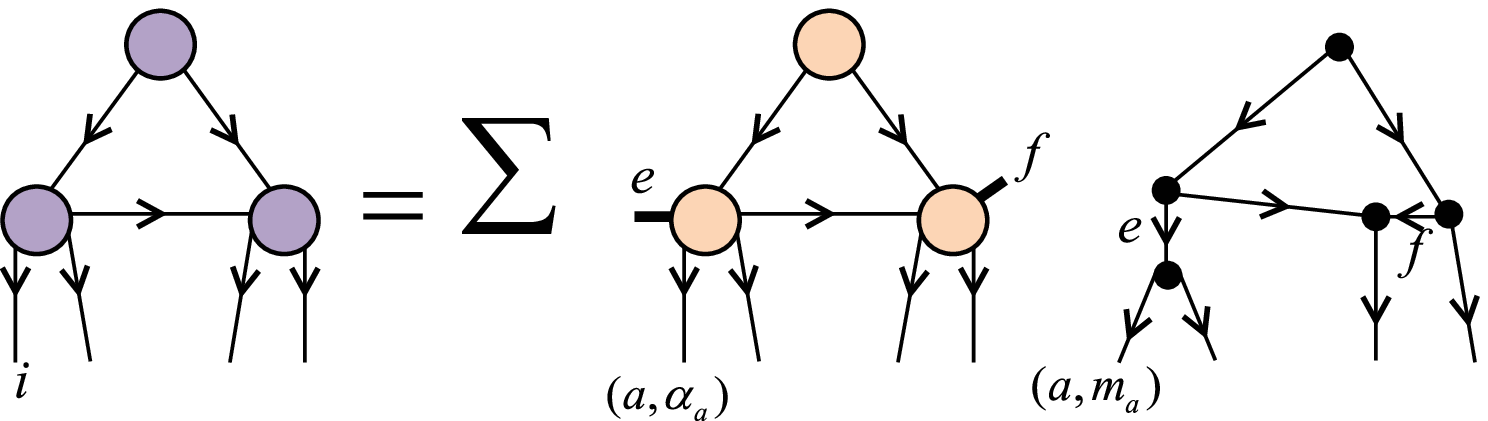}
\caption{(Color online) A TN for a symmetric state $\ket{\Psi}\in \mathbb{V}^{\otimes N}$ of lattice $\mathcal{L}$ (Fig. \ref{fig:symmTN}) is expressed as a linear superposition of spin networks. The sum runs over the intermediate indices that carry charges $e$ and $f$ (shown explicitly) as well as all indices shared by two tensors.}
\label{fig:spinNetwork}
\end{figure}%
%%%%%%%%%%%%%%%%%%%%%%%%%%%%%%%%%%%%%
%%%%%%%%%%%%%%%%%%%%%%%%%%%%%%%%%%%%%

We can now investigate how the TN decomposes if we write each of its tensors $T$ in the $(P,Q)$ form of Eq. \ref{eq:decoT} (see Fig. \ref{fig:spinNetwork}). For any fixed value of all the charges, the whole TN factorizes into two terms. The first one is a TN of degeneracy tensors. The second one is a directed graph with edges labeled by irreps of $\mathcal{G}$ and vertices labeled by intertwining operators. This is nothing other than a \emph{spin network} \cite{SpinNetwork}, a well-known object in mathematical physics and, especially, in loop quantum gravity \cite{LQG}, where it is used to describe states of quantum geometry. Accordingly, a symmetric TN for the state $\ket{\Psi}\in \mathbb{V}^{\otimes N}$ of a lattice $\mathcal{L}$ of $N$ sites can be regarded as a linear superposition of spin networks with $N$ open edges. The number of spin networks in the linear superposition grows exponentially with the size of the TN. The expansion coefficients are given by the degeneracy tensors.

Computationally, the present characterization of a symmetric TN is of interest for several reasons. First of all, it allows us to describe a state $\ket{\Psi}^{\otimes N}$ with \emph{specific} quantum numbers, which are preserved exactly during approximate numerical simulations. Let us consider as an example the group U(1), with charge $n$ corresponding to particle number ($n=0,\pm 1,\pm 2, \ldots$), and the group SU(2), with charge $j$ corresponding to the spin ($j=0,1/2,1,3/2, \ldots$). The symmetric TN can be used to describe a state with e.g. zero particles ($n=0$) and zero spin ($j=0$) respectively; or, more generally, covariant states with any value of $n$ and $j$ \cite{covariant}. 

Secondly, the ($P,Q$)-decomposition (\ref{eq:decoT}) concentrates all the degrees of freedom of a symmetric tensor $T$ in the degeneracy tensors $P$, producing a more compact description. For instance, for the U(1) and SU(2) groups, an approximation 
of the ground state of the antiferromagnetic Heisenberg spin-$\frac{1}{2}$ chain with a MERA of bond dimension $\chi=21$ requires $5$ and $35$ times less parameters than with non-symmetric tensors respectively \cite{U1,SU2}. 

%%%%%%%%%%%%%%%%%%%%%%%%%%%%%%%%%%%%%
%%%%%%%%%%%%%%%%%%%%%%%%%%%%%%%%%%%%%
\begin{figure}[t]
  \includegraphics[width=7.48cm]{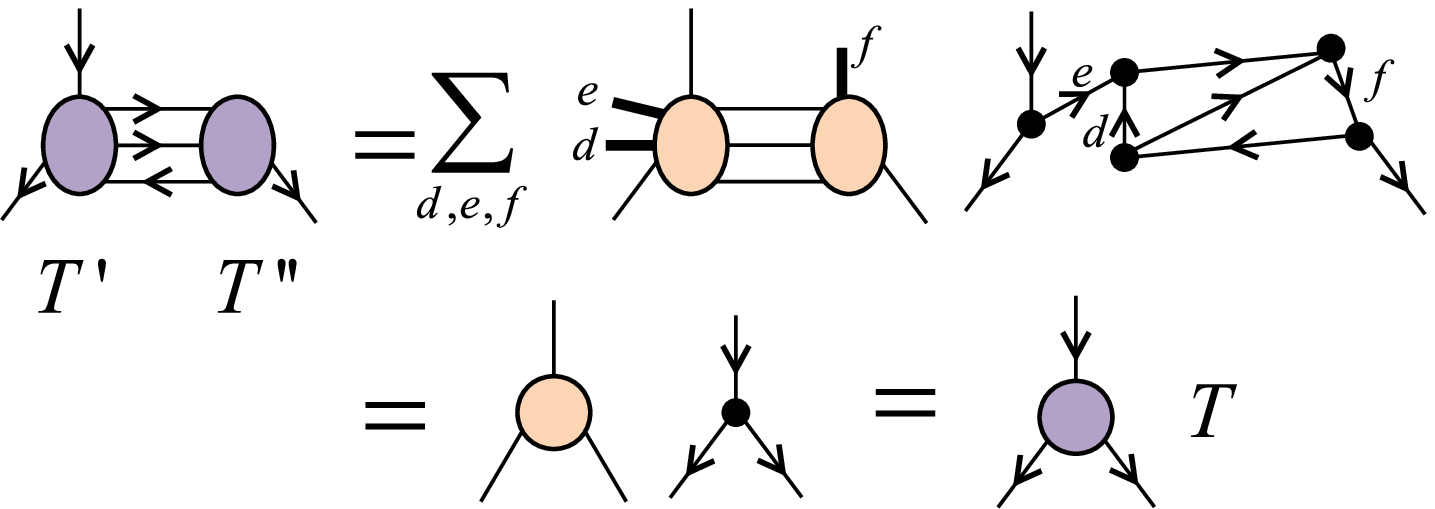}
\caption{(Color online) Product of two symmetric tensors. Only the intermediate charges $d$, $e$ and $f$ are explicitly shown. Additional sums apply to all indices shared by two tensors. The computation involves evaluating spin networks.}
\label{fig:T1T2}
\end{figure}%
%%%%%%%%%%%%%%%%%%%%%%%%%%%%%%%%%%%%%
%%%%%%%%%%%%%%%%%%%%%%%%%%%%%%%%%%%%%

In addition, the ($P,Q$)-decomposition (\ref{eq:decoT}) lowers the cost of simulations significantly. Consider the multiplication of two tensors (Fig. \ref{fig:T1T2}) which is central to most TN algorithms. Cost reductions come from two fronts: \\
(i) \emph{Block-sparse matrices.---} The most costly step in multiplying two tensors $T'$ and $T''$ consists of multiplying two matrices $M'$ and $M''$ obtained from $T'$ and $T''$. These matrices are of the form of Eq. \ref{eq:Schur}, and therefore their multiplication can be done blockwise, 
\begin{equation}
M = M'M''= \bigoplus_{a} \left[ (M'^a M''^a) \otimes \tilde{\mathbb{I}}^a\right].
\end{equation} 
(ii) \emph{Pre-computation.---} Given a $(P,Q)$-decomposition of tensor $T$, another ($\tilde{P},\tilde{Q}$)-decomposition (as required e.g. to obtain matrices $M'$ and $M''$ above) involves a linear map $\Gamma$, 
\begin{equation}
\tilde{P} = \Gamma(P).
\label{eq:Gamma}	
\end{equation}
This map $\Gamma$, of which Eq. \ref{eq:Fmove2} is an example, is completely determined by the symmetry. In those TN algorithms that proceed by repeating a sequence of manipulations, map $\Gamma$ can be %pre-
computed once and stored in memory for repeated usage.

A more detailed explanation of algorithmic details, as well as practical examples of the gains obtained using invariant tensors, is presented in Refs. \cite{U1} and \cite{SU2} for the groups U(1) and SU(2), respectively. Ref. \cite{MERAanti} exploited the U(1) symmetry in a 2D MERA calculation that involved tensors with up to twelve indices.

Finally, the connection between symmetric TNs and spin networks allow us to import into the context of TN algorithms techniques developed to evaluate spin networks in loop quantum gravity. Such techniques can be used e.g. to compute the linear map $\Gamma$ of Eq. \ref{eq:Gamma}. Conversely, TN algorithms may also prove useful in loop quantum gravity, since they allow e.g. %to efficiently manipulate 
the efficient manipulation of
superpositions of an exponentially large number of spin networks. 

We thank Ian McCulloch for continued discussions. Support from the Australian Research Council (APA, FF0668731, DP0878830) is acknowledged.

%%%%%%%%%%%%%%%%%%%%%%%%%%%%%%%%%%%%%%%%%%%%%%%%%%%%%%%%%%%%

%

\begin{thebibliography}{13}
%
\bibitem{MPS} S. Ostlund and S. Rommer, Phys. Rev. Lett. \tbf{75}, 3537 (1995).
M. Fannes, B. Nachtergaele, and R. Werner, Commun. Math. Phys. \tbf{144}, 443 (1992).
%
\bibitem{PEPS} 
F. Verstraete and J. I. Cirac, arXiv:cond-mat/0407066v1. 
G. Sierra and M.A. Martin-Delgado, arXiv:cond-mat/9811170v3. 
%T. Nishino, K. Okunishi, J. Phys. Soc. Jpn. 67, 3066, 1998. %,arXiv:cond-mat/9804134v2.
%
\bibitem{MERA} G. Vidal, Phys. Rev. Lett. \tbf{99}, 220405 (2007). G. Vidal, Phys. Rev. Lett. \tbf{101}, 110501 (2008).  
%
\bibitem{MERAanti}
G. Evenbly and G. Vidal,  Phys. Rev. Lett. \tbf{104}, 187203 (2010).
%
\bibitem{SpinNetwork} R. Penrose, \emph{Angular momentum: an approach to combinatorial space-time}, 1971. (http://math.ucr.edu/home/ baez/penrose/).
%
\bibitem{LQG}
C. Rovelli and L. Smolin, Physical Review D \tbf{52}, 5743 (1995). %; gr-qc/9505006
C. Rovelli, Living Reviews in Relativity (http://www.livingreviews.org/lrr-2008-5).
%
\bibitem{symmMPS}  
S. R. White, Phys. Rev. Lett. \tbf{69}, 2863 (1992);
S.~Ostlund and S. Rommer, Phys. Rev. Lett. \tbf{75}, 3537 (1995). %,  arXiv:cond-mat/9503107v1
G.~Sierra and T. Nishino, Nucl. Phys. \tbf{B495}, 505 (1997). %  arXiv:cond-mat/9610221v1. 
I. McCulloch and M. Gulacsi, Europhys. Lett. \tbf{57}, 852 (2002). %, arXiv:cond-mat/0012319.
I. McCulloch, J. Stat. Mech. (2007) P10014. %, arXiv:cond-mat/0701428.
S.~Singh, H.-Q. Zhou, and G. Vidal, New J. Phys. \tbf{12} (2010) 033029. %, arXiv:cond-mat/0701427. 
D. Perez-Garcia et al., Phys. Rev. Lett. \tbf{100}, 167202 (2008). %, arXiv:0802.0447v1. 
M. Sanz et al., Phys. Rev. A \tbf{79}, 042308 (2009). %, arXiv:0901.2223v1 
%
\bibitem{covariant} A set of states $\ket{\Psi_{t}}$ that transform \textit{covariantly}, $(U_g)^{\otimes N} \ket{\Psi_t} = \sum_{t'} (W_g)_{tt'}\ket{\Psi_{t'}}$, where $W$ is a unitary representation of $\mathcal{G}$, can be represented by an invariant pure state $\ket{\Phi} \propto \sum_t \ket{\Psi_t}\ket{t}$ of lattice $\mathcal{L}$ and one additional site on which the group acts with $W_g^{\dagger}$. 
%The same is true for an invariant mixed state $\rho \propto \sum_t \proj{\Psi_t}$, with $(U_g)^{\otimes N} \rho (U_g^{\dagger})^{\otimes N}$.
%
\bibitem{group} J. F. Cornwell, {\it Group Theory in Physics}, (Academic Press, 1997).
%
\bibitem{degeneracy} In non-multiplicity free groups, such as SU(3), where $N_{ab}^c$ might be larger than 1, the coupled basis $\ket{c,o_c,\mu}$ and tensor $S^{abc}_{\mu}$ must include an extra index $\mu=1,\ldots,N_{ab}^c$. See, for example, R. N. C. Pfeifer et al., Phys. Rev. B \tbf{82}, 115126 (2010).
%
\bibitem{abelian} When $\mathcal{G}$ is an Abelian group, such as U(1), the tensor product $\mathbb{V}^{a}\otimes \mathbb{V}^{b}$ of two irreps only gives rise to one irrep $\mathbb{V}^{c}$, so that no intermediate charges $e_1, e_2, \ldots, e_{t'}$ need to be specified in Eq. \ref{eq:decoT}, simplifying significantly the decomposition of symmetric tensors.
%
\bibitem{U1} S. Singh, R. Pfeifer, and G. Vidal, arXiv:1008.4774v1. 
\bibitem{SU2}S. Singh and G. Vidal, in preparation.
%
\end{thebibliography}
\end{document}